\documentclass{bcp}
\usepackage{amsmath,amsthm}
\usepackage{amssymb}
\usepackage{color}

\usepackage[T1]{fontenc}

\input xy
\xyoption{all} \tolerance=500

\newtheorem{theorem}{Theorem}[section]

\theoremstyle{definition}
\newtheorem{definition}[theorem]{Definition}
\newtheorem{example}[theorem]{Example}
\newtheorem{remark}[theorem]{Remark}

\font\black=cmbx10 \font\sblack=cmbx7 \font\ssblack=cmbx5 \font\blackital=cmmib10  \skewchar\blackital='177
\font\sblackital=cmmib7 \skewchar\sblackital='177 \font\ssblackital=cmmib5 \skewchar\ssblackital='177
\font\sanss=cmss10 \font\ssanss=cmss8 
\font\sssanss=cmss8 scaled 600 \font\blackboard=msbm10 \font\sblackboard=msbm7 \font\ssblackboard=msbm5
\font\caligr=eusm10 \font\scaligr=eusm7 \font\sscaligr=eusm5  \font\fraktur=eufm10
\font\sfraktur=eufm7 \font\ssfraktur=eufm5

\font\bsymb=cmsy10 scaled\magstep2
\def\all#1{\setbox0=\hbox{\lower1.5pt\hbox{\bsymb
       \char"38}}\setbox1=\hbox{$_{#1}$} \box0\lower2pt\box1\;}
\def\exi#1{\setbox0=\hbox{\lower1.5pt\hbox{\bsymb \char"39}}
       \setbox1=\hbox{$_{#1}$} \box0\lower2pt\box1\;}

\def\tx#1{{\fam0\relax#1}}

\newfam\bifam
\textfont\bifam=\blackital \scriptfont\bifam=\sblackital \scriptscriptfont\bifam=\ssblackital

\newfam\blfam
\textfont\blfam=\black \scriptfont\blfam=\sblack \scriptscriptfont\blfam=\ssblack

\newfam\bbfam
\textfont\bbfam=\blackboard \scriptfont\bbfam=\sblackboard \scriptscriptfont\bbfam=\ssblackboard

\newfam\ssfam
\textfont\ssfam=\sanss \scriptfont\ssfam=\ssanss \scriptscriptfont\ssfam=\sssanss
\def\sss#1{{\fam\ssfam\relax#1}}

\newfam\clfam
\textfont\clfam=\caligr \scriptfont\clfam=\scaligr \scriptscriptfont\clfam=\sscaligr

\newfam\frfam
\textfont\frfam=\fraktur \scriptfont\frfam=\sfraktur \scriptscriptfont\frfam=\ssfraktur

\def\hpb#1{\setbox0=\hbox{${#1}$}
    \copy0 \kern-\wd0 \kern.2pt \box0}
\def\vpb#1{\setbox0=\hbox{${#1}$}
    \copy0 \kern-\wd0 \raise.08pt \box0}

\def\pmb#1{\setbox0\hbox{${#1}$} \copy0 \kern-\wd0 \kern.2pt \box0}
\def\pmbb#1{\setbox0\hbox{${#1}$} \copy0 \kern-\wd0
      \kern.2pt \copy0 \kern-\wd0 \kern.2pt \box0}
\def\pmbbb#1{\setbox0\hbox{${#1}$} \copy0 \kern-\wd0
      \kern.2pt \copy0 \kern-\wd0 \kern.2pt
    \copy0 \kern-\wd0 \kern.2pt \box0}
\def\pmxb#1{\setbox0\hbox{${#1}$} \copy0 \kern-\wd0
      \kern.2pt \copy0 \kern-\wd0 \kern.2pt
      \copy0 \kern-\wd0 \kern.2pt \copy0 \kern-\wd0 \kern.2pt \box0}
\def\pmxbb#1{\setbox0\hbox{${#1}$} \copy0 \kern-\wd0 \kern.2pt
      \copy0 \kern-\wd0 \kern.2pt
      \copy0 \kern-\wd0 \kern.2pt \copy0 \kern-\wd0 \kern.2pt
      \copy0 \kern-\wd0 \kern.2pt \box0}

\mathchardef\za="710B  
\mathchardef\zb="710C  
\mathchardef\zg="710D  
\mathchardef\zd="710E  
\mathchardef\zve="710F 
\mathchardef\zz="7110  
\mathchardef\zh="7111  
\mathchardef\zvy="7112 
\mathchardef\zi="7113  
\mathchardef\zk="7114  
\mathchardef\zl="7115  
\mathchardef\zm="7116  
\mathchardef\zn="7117  
\mathchardef\zx="7118  
\mathchardef\zp="7119  
\mathchardef\zr="711A  
\mathchardef\zs="711B  
\mathchardef\zt="711C  
\mathchardef\zu="711D  
\mathchardef\zvf="711E 
\mathchardef\zq="711F  
\mathchardef\zc="7120  
\mathchardef\zw="7121  
\mathchardef\ze="7122  
\mathchardef\zy="7123  
\mathchardef\zf="7124  
\mathchardef\zvr="7125 
\mathchardef\zvs="7126 
\mathchardef\zf="7127  
\mathchardef\zG="7000  
\mathchardef\zD="7001  
\mathchardef\zY="7002  
\mathchardef\zL="7003  
\mathchardef\zX="7004  
\mathchardef\zP="7005  
\mathchardef\zS="7006  
\mathchardef\zU="7007  
\mathchardef\zF="7008  
\mathchardef\zW="700A  
\mathchardef\zC="7009  

\newcommand{\be}{\begin{equation}}
\newcommand{\ee}{\end{equation}}

\newcommand{\bea}{\begin{eqnarray}}
\newcommand{\eea}{\end{eqnarray}}
\newcommand{\beas}{\begin{eqnarray*}}
\newcommand{\eeas}{\end{eqnarray*}}
\def\*{{\textstyle *}}
\newcommand{\R}{{\mathbb R}}

\newcommand{\GL}{\mathrm{GL}}
\def\ssT{\sss T}

\newcommand{\we}{\wedge}

\newcommand{\ot}{\otimes}
\newcommand{\s}{{\textstyle *}}

\newcommand{\pa}{\partial}
\newcommand{\ti}{\times}

\newcommand{\ad}{{\rm ad}}

\def\ran{\rangle}

\def\cD{{\cal D}}

\def\cR{{\cal R}}

\def\wh{\widehat}

\def\la{\langle}
\def\ran{\rangle}

\def\sJ{{\sss J}}

\def\sP{{\sss P}}

\def\sT{{\sss T}}

\def\st{{\sss t}}

\def\sj{{\sss j}}
\def\sq{{\sss q}}

\def\xd{\tx{d}\,}
\def\xi{\tx{i}}

\def\cD{\cal D}

\def\eza{\za}
\def\ezb{\zb}

\def\xd{\operatorname{d}\!}

\def\s*{{\scriptstyle *}}

\def\Mi{\mathrm{Mi}}
\def\uxd{{\underline{\mathrm{d}}}}
\def\g{{g}}

\def\rel{-\!\!-\triangleright}
\def\rel{{-\!\!\!-\!\!\rhd}}
\newdir{|>}{%
!/4.5pt/@{|}*:(1,-.2)@^{>}*:(1,+.2)@_{>}}
\newdir{ (}{{}*!/-5pt/@^{(}}

\begin{document}

\mathclass{Primary 70H50; Secondary 53D17, 70H50, 83E30.}

\abbrevauthors{J. Grabowski}
\abbrevtitle{New developments...}

\title{New developments in geometric mechanics}

\author{Janusz Grabowski}
\address{Institute of Mathematics, Polish Academy of Sciences\\
\'Sniadeckich 8, 00-956 Warszawa, Poland\\
E-mail: jagrab@impan.pl}

\author{Andrew J. Bruce}
\address{Institute of Mathematics, Polish Academy of Sciences\\
\'Sniadeckich 8, 00-956 Warszawa, Poland\\
E-mail: andrew@impan.pl}

\author{Katarzyna Grabowska}
\address{Department of Physics, University of Warsaw\\
Pasteura 5, 02-093 Warszawa, Poland\\
E-mail: konieczn@fuw.edu.pl}

\author{Pawe{\l} Urba\'nski}
\address{Department of Physics, University of Warsaw\\
Pasteura 5, 02-093 Warszawa, Poland\\
E-mail: konieczn@fuw.edu.pl}

\maketitlebcp

\begin{abstract}
We review the concept of a graded bundle, which is a generalisation of a vector bundle, its linearisation, and a double structure of this kind. We then present applications of these structures in geometric mechanics  including systems with higher order Lagrangian and the Plateau problem.
\end{abstract}

\section{Introduction}
In this paper we report on the key results of our collective works \cite{BGG1, BGG2, GGU}. The broad idea of these works can be stated as \emph{applying the notion of  graded bundles  to the setting of geometric mechanics in the spirit of W.M.~Tulczyjew}.

Recall that Grabowski and Rotkiewicz \cite{GR} established the one-to-one correspondence between manifolds that admit non-negatively graded local coordinates and manifolds equipped with an action of the monoid of multiplicative reals, or \emph{homogeneity structures} in the language developed in that paper. Such manifolds are  referred to as \emph{graded bundles} for reasons that we will shortly explain.  The cardinal examples of graded bundles are the higher order tangent bundles, which of course play a central r\^{o}le in  (standard) higher order mechanics.

We  describe a quite general geometric set-up of higher order mechanics for which the (higher order) velocities get replaced with elements of a graded bundle. To realise this we employ the notion of a \emph{weighted Lie algebroid} \cite{BGG2}. Such manifolds have simultaneously the structure of a graded bundle and a Lie algebroid that are compatible in a precise sense. The  approach we develop  makes use of first order mechanics on Lie algebroids subject to affine vakonomic constraints and, as a result, we are lead to consider relations as opposed to genuine maps. The higher order flavour  is due to the fact that  underlying a weighted Lie algebroid is a graded bundle: there is an associated series of affine fibrations that mimic higher order tangent bundles. The standard  description of higher order Lagrangian mechanics can naturally be accommodated within this framework, as can higher order mechanics on a Lie algebroid.  For the case of higher order mechanics on a Lie algebroid, the Euler--Lagrange equations we obtain agree  with the ones obtained by {Colombo \& de Diego} \cite{CD}, {J\'{o}\'{z}wikowski \&   Rotkiewicz} \cite{JR1,JR2} and {Mart\'{\i}nez} \cite{Ma1}. One should note that our approach is completely geometric and we do not employ any standard tools from the calculus of variations (like variations etc.). The Hamiltonian formalism can also be described within this framework, however we will not discuss this in this paper and refer the reader to \cite{BGG1}.

 The challenge of describing mechanical systems configured on Lie groupoids and their reduction to Lie algebroids was first posted by Weinstein \cite{Weinstein:1996}. This challenge was taken up by many authors and various approaches developed,  a rather incomplete list  is given in \cite{BGG1}.  The notion of the \emph{Tulczyjew triple} for a Lie algebroid, as we shall understand it, was first given in \cite{Grabowska:2006}. It was based on a framework for Lagrangian and Hamiltonian formalisms developed by Tulczyjew \cite{Tulczyjew:1976,Tulczyjew:1976a,Tulczyjew:1989} and a corresponding description of Lie algebroids \cite{GU3,GU1}. The motivation for extending the geometric tools of the Lagrangian formalism on tangent bundles to Lie algebroids comes from the  fact that reductions usually push one out of the environment of tangent bundles and into the world of Lie algebroids. In a similar way, reductions of higher order tangent bundles, which is where higher order mechanical Lagrangians ``live'', will push one into the environment of ``higher Lie algebroids''. Weighted Lie algebroids turn out to give a clear geometric framework for reductions of higher order tangent bundles.

Graded bundles play also important r\^ole in our approach to the  geometric mechanics of (classical bosonic) \emph{strings} \cite{GGU}.   The constructions  cover the case of higher dimensional \emph{branes}, though for simplicity we  only discuss strings in any detail.  The basic tools for point-particles is not adequate for the description of strings in geometric mechanics. The geometric approach to classical strings as presented here is based on morphisms of double graded bundles and the  (canonical) multisymplectic structure on $\wedge^{2} \sT^{*}M$. Again, we have a  generalisation of the Tulczyjew triple, now adapted for the setting of classical strings.  The phase space, the phase equations and Legendre transformation relating the Lagrangian and Hamiltonian pictures  are obtained  in a purely geometric way. The Euler--Lagrange equations for strings in this geometric set-up are of course the equations of motion for the Nambu--Goto string.


\section{Graded and double graded bundles}

A \emph{vector bundle} is a locally trivial fibration $\zt:E\to M$ which, locally over $U\subset M$, reads $\zt^{-1}(U)\simeq U\ti\R^n$ and admits an atlas in which local trivializations transform linearly in fibers
$$U\cap V\ti\R^n\ni(x,y)\mapsto(x,A(x)y)\in U\cap V\ti\R^n\,,$$
$A(x)\in\GL(n,\R)$.

The latter property can also be expressed in the terms of the gradation in which base coordinates $x$ have degrees $0$ and `linear coordinates' $y$ have degree $1$. Linearity in $y's$ is now equivalent to the fact that changes of coordinates respect the degrees. Morphisms in the category of vector bundles are represented by commutative diagram of smooth maps
$$\xymatrix{
E_1\ar[rr]^{\Phi} \ar[d]^{\zt_1} && E_2\ar[d]^{{\zt_2}} \\
M_1\ar[rr]^{{\varphi}} && M_2 }
$$
being linear in fibres. As canonical examples and constructions can serve e.g. $\sT M$, $\sT^*M$, $E\ot_MF$, $\we^kE$, etc.

A natural generalization of the vector bundle is the concept of a \emph{graded bundle} $\zt:F\to M$ (cf. \cite{GR}). We have a local trivialization by $U\ti\R^n$ as before, but with the difference that the local coordinates $(y^1,\dots,y^n)$ in the fibres have now associated positive integer \emph{weights} (or \emph{degrees}) $w_1,\dots,w_n$, that are preserved by changes of local trivializations:
$$U\cap V\ti\R^n\ni(x,y)\mapsto(x,A(x,y))\in U\cap V\ti\R^n\,.$$

One can show that in this case $A(x,y)$ must be polynomial in fiber coordinates, i.e. any graded bundle is a \emph{polynomial bundle} \cite{GR}.
As these polynomials need not to be linear, graded bundles do not have, in general, vector space structure in fibers. For instance, if $(y,z)\in\R^2$ are coordinates of degrees $1,2$, respectively, then the map $(y,z)\mapsto (y,z+y^2)$ is a diffeomorphism preserving the degrees, but it is nonlinear.

If all $w_i\le r$, we say that the graded bundle is \emph{of degree} $r$.
In the above terminology, vector bundles are just graded bundles of degree $1$.
Graded bundles $F_k$ of degree $k$ admit, like many jet bundles, a tower of affine fibrations by the bundles of lower degrees
$$
F_{k} \stackrel{\tau^{k}}{\longrightarrow} F_{k-1} \stackrel{\tau^{k-1}}{\longrightarrow}   \cdots \stackrel{\tau^{3}}{\longrightarrow} F_{2} \stackrel{\tau^{2}}{\longrightarrow}F_{1} \stackrel{\tau^{1}}{\longrightarrow} F_{0} = M\,.
$$

\begin{example}
\emph{Higher tangent bundles} $\sT^kM$, with canonical coordinates $(x,\dot x,\ddot x,\dddot x,\dots)$ of degrees, respectively, $0,1,2,3,$ etc. are canonical examples of graded bundles.
\end{example}
\begin{example} If $\zt:E\to M$ is a vector bundle, then $\we^r\sT E$ is canonically a graded bundle of degree $r$ with respect to the projection
    $$\we^r\sT\zt:\we^r\sT E\to \we^r\sT M\,.$$
\end{example}
\begin{remark}Note that similar objects has been used in supergeometry by  Voronov \cite{Voronov:2001qf} under the name of \emph{non-negatively graded (super)manifolds}. If the Grassmann parity of the coordinates is given by the weight of the coordinates (mod 2) then the resulting supermanifolds are known as \emph{N-manifolds}  following  \v{S}evera \cite{Severa:2005} and Roytenberg \cite{Roytenberg:2001}. However, we will work with classical purely even manifolds only.
\end{remark}

With the use of coordinates $(x^\za,y^a)$ with degrees $0$ for basic coordinates $x^\za$, and degrees $w_a>0$ for the fibre coordinates $y^a$, we can define on the graded bundle $F$ a globally defined \emph{weight vector field} (\emph{Euler vector field})
$$\nabla_F=\sum_aw_ay^a\pa_{y^a}\,.$$

The flow of the weight vector field extends to a smooth action $\R\ni t\mapsto h_t$ of multiplicative reals on $F$, $h_t(x^\zm,y^a)=(x^\zm,t^{w_a}y^a)$. Such an action $h:\R\ti F\to F$, $h_t\circ h_s=h_{ts}$, we will call a \emph{homogeneity structure} \cite{GR}.
\begin{definition}
A function $f:F\to\R$ is called \emph{homogeneous of degree (weight) $k$} if $(h_t^*f)(x):=f(h_t(x))=t^kf(x)$; similarly for the homogeneity of tensor fields.
\emph{Morphisms} of two homogeneity structures $(F_i,h^i)$, $i=1,2$, are smooth maps $\Phi:F_1\to F_2$ intertwining the $\R$-actions: $\Phi\circ h^1_t=h^2_t\circ\Phi$. Consequently, a \emph{homogeneity substructure} is a smooth submanifold $S$ invariant with respect to $h$, $h_t(S)\subset S$.
\end{definition}
\subsection{Double graded bundles}

The fundamental fact (cf. \cite{GR}) says that graded bundles and homogeneity structures are in fact equivalent concepts.

\begin{theorem}
For any homogeneity structure $h$ on a manifold $F$, there is a smooth submanifold $M=h_0(F)\subset F$, a non-negative integer $k\in\mathbb N$, and an $\R$-equivariant map
$\Phi_h^k:F\to \sT^kF_{|M}$
which identifies $F$ with a graded submanifold of the graded bundle $\sT^kF$. In particular, there is an atlas on $F$ consisting of local homogeneous functions.
\end{theorem}

As two graded bundle structure on the same manifold are just two homogeneity structures, the obvious concept of compatibility leads to the following \cite{GR}.
\begin{definition} A \emph{double graded bundle} is a manifold equipped with two homogeneity structures $h^1,h^2$ which are \emph{compatible} in the sense that
$$h^1_t\circ h^2_s=h^2_s\circ h^1_t\quad \text {for all\ } s,t\in\R\,.$$
\end{definition}
This covers of course the concept of a \emph{double vector bundle} of Pradines \cite{P}
and extends to \emph{$n$-tuple} graded bundles in the obvious way.

\subsection{Lifts} If $\zt:F\to M$ is a graded bundle of degree $k$, then $\sT F$ and $\sT^*F$ carry canonical double graded bundle structure: one is the obvious vector bundle, the other is of degree $k$. A double graded bundle whose one structure is linear we will call a \emph{$\mathcal{GL}$-bundle}. There are also lifts of graded structures on $F$ to $\sT^r F$.

In particular, if $\zt:E\to M$ is a vector bundle, then $\sT E$ and $\sT^*E$ are double vector bundles. The latter is isomorphic with $\sT^*E^*$.
As a linear Poisson structure on $E^*$ yields a map $\sT^*E^*\to\sT E^*$, a Lie algebroid structure on $E$ can be encoded as a morphism of double vector bundles, $\ze:\sT^*E\to\sT E^*$ (see \cite{GU3,GU1}).

\begin{example}
If $\zt:E\to M$ is a vector bundle, then $\we^k\sT E$ is canonically a $\mathcal{GL}$-bundle:
$$
        {\xymatrix@R-7mm @C-4mm{ & \wedge^k\sT E \ar[ld] \ar[rd] & \cr
        \quad E \ar[rd] & & \wedge^k\sT M \ar[ld]  \cr & M  & }}\,.
$$
\end{example}

\section{Tulczyjew triples}
The canonical symplectic form $\zw_M$ on $\sT^\* M$ induces an isomorphism
$$\zb_M:\sT\sT^\* M\to\sT^\*\sT^\* M\,.$$
Composing it with $\cR_{\sT M}$, where
$$\cR_E:\sT^\*E^\*\to \sT^\*E$$
is the well-known canonical isomorphism (see e.g. \cite{KU,Mc,Ur}),
 we get the map
$$\za_M:\sT\sT^\* M\to\sT^\*\sT M\,.$$
Using the standard coordinates $(x^\zm,\dot{x}^\zn)$ and $(x^\zm,p_\zn)$ on $\sT M$ and $\sT^*M$, respectively, and the adapted coordinates on $\sT^*\sT M$ and $\sT\sT^*M$, we can write
\be\label{alpha} \za(x,p,\dot x,\dot p)=(x,\dot x,\dot p,p)\,.
\ee
This gives rise to the commutative diagram of \emph{double vector bundle (iso)morphisms} (Tulczyjew triple)
\be\label{tt}
    \xymatrix@R-4mm @C-10mm
        { & \sT^\*\sT^\*M \ar[ldd]_*{} \ar[rd]^*{} & & & \sT \sT^\* M \ar[rrr]^*{{\eza}_M}
        \ar[lll]_*{\ezb_M} \ar[ldd]^*{} \ar[rd]^*{}& & & \sT^\*\sT M \ar[ldd]^*{} \ar[rd]^*{} & \cr
        & & \sT M \ar[ldd]^*{} & & & \sT M  \ar[ldd]^*{} \ar[lll]^*{} \ar[rrr]^*{} & & & \sT M \ar[ldd]^*{}  \cr
         \sT^\* M  \ar[rd]^*{}  & & & \sT^\* M \ar[rrr]^*{} \ar[lll]^*{} \ar[rd]^*{}& & & \sT^\* M
        \ar[rd]^*{} & & \cr
        & M  & & &  M \ar[rrr]^*{} \ar[lll]^*{} & & & M & }.
\ee

Note that the mapping $\za_M$ can be obtained directly as the dual to the `canonical flip' $\zk_M:\sT\sT M\to \sT\sT M$
 which is an isomorphism of two vector bundle structures on $\sT\sT M$:
\be\label{kappa}
    {\xymatrix@R-3mm @C-2mm{  & \sT \sT M \ar[ldd]_*{\sT\zt_M} \ar[rd]^*{\zt_{\ssT M}}
    \ar[rrr]^*{\zk_M} & & & \sT \sT M \ar[ldd]_(.3)*{\zt_{\ssT M }}  \ar[rd]^*{\sT\zt_M} & \cr
    & & \sT M \ar[ldd]_(.3)*{\zt_M}  \ar[rrr]^(.7)*{\text{id}} & & & \sT M \ar[ldd]_*{\zt_M} \cr
    \sT M  \ar[rd]^*{\zt_M} \ar[rrr]^(.7)*{\text{id} } & & &
    \sT M \ar[rd]^*{\zt_M}  & &  \cr
    & M \ar[rrr]^*{\text{id}} & & & M &}} .\ \ \
\ee
Indeed, the duals of these two vector bundle structures on $\sT\sT M$ are $\sT^\*\sT M$ and $\sT\sT^\* M$, and $\za_M$ can be understood as the dual map of $\zk_M$.

\medskip\noindent
The map $\zk_M$, as well as $\za_M$ and $\zb_M$, encodes the Lie algebroid structure of $\sT M$ and note that no brackets are needed (cf. \cite{GU3,GU1}).

The Lagrangian and Hamiltonian formalisms have simple description in terms of the Tulczyjew triple. The true physical dynamics, the \emph{phase dynamics}, will be described as an implicit first order differential equation on the \emph{phase space} $\sT^*M$, given by a submanifold $\mathcal{D}\subset\sT\sT^*M$. Note that a solution of an implicit differential equation $\mathcal{D}\subset\sT N$ is a curve $\zg:\R\to N$ such that its \emph{tangent prolongation} $\st\zg:\R\to\sT N$ takes values in $\cD$.

\subsection{The Tulczyjew triple - Lagrangian side}

Denote with $M$ positions of our system, with $\sT M$ (kinematic) configurations, and let $L:\sT M\rightarrow \R$ be a Lagrangian function. We have the diagram

$$
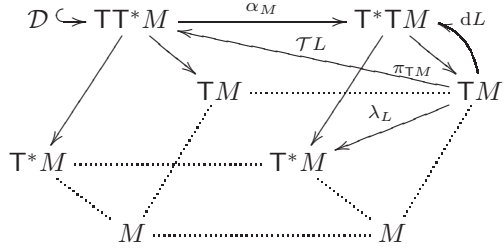
\xymatrix@C-20pt@R-10pt{
{\mathcal{D}}\ar@{ (->}[r]& \sT\sT^\ast M \ar[rrr]^{\alpha_M} \ar[dr]\ar[ddl]
 & & & \sT^\ast\sT M\ar[dr]_{\pi_{\sT M}}\ar[ddl] & \\
 & & \sT M\ar@{.}[rrr]\ar@{.}[ddl]
 & & & \sT M \ar@{.}[ddl]\ar@/_1pc/[ul]_{\xd L}\ar[dll]_{\zl_L}\ar[ullll]_{\mathcal{T}L}\\
 \sT^\ast M\ar@{.}[rrr]\ar@{.}[dr]
 & & & \sT^\ast M\ar@{.}[dr] & &  \\
 & M\ar@{.}[rrr]& & & M &
}$$
The dynamics
$$\mathcal{D}=\alpha_M^{-1}(\xd L(\sT M)))=\mathcal{T}L(\sT M)\,,$$
is given as the range of the \emph{Tulczyjew differential} $\mathcal{T}L=\za_M^{-1}\circ\xd L$.
In local coordinates,
$$\mathcal{D}=\left\{(x,p,\dot x,\dot p):\;\; p=\frac{\partial L}{\partial \dot x},\quad \dot p=\frac{\partial L}{\partial x}\right\}\,.$$
Notice in the diagram the \emph{Legendre map},
$$\zl_L:\sT M\rightarrow \sT^\ast M, \;\; \quad \zl_L(x,\dot x)=
(x,\frac{\partial L}{\partial \dot x})\,.$$

\subsection{The Tulczyjew triple - Hamiltonian side}
The Hamiltonian formalism looks analogously. If {$H:\sT^\ast M\rightarrow \R$} is a Hamiltonian function, from the Hamiltonian side of the triple
{$$\hskip-1.2cm
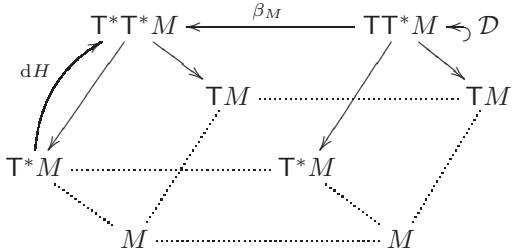
\xymatrix@C-20pt@R-10pt{
 & \sT^\ast\sT^\ast M  \ar[dr] \ar[ddl]
 & & & \sT\sT^\ast M\ar[dr]\ar[ddl] \ar[lll]_{\beta_M}&
 { \mathcal{D}}\ar@{ (->}[l] \\
 & & \sT M\ar@{.}[rrr]\ar@{.}[ddl]
 & & & \sT M \ar@{.}[ddl]\\
 \sT^\ast M\ar@{.}[rrr]\ar@{.}[dr] \ar@/^1pc/[uur]^{\xd H}
 & & & \sT^\ast M\ar@{.}[dr] & &  \\
 & M\ar@{.}[rrr]& & & M &
}\qquad$$}
we derive the phase dynamics in the form
{$$\mathcal{D}=\beta_M^{-1}(\xd H(\sT^\ast M))\,.$$}
It is automatically explicit, i.e. generated by the corresponding Hamiltonian vector field, so corresponds to a phase dynamics induced by a Lagrangian function only in regular cases.
In local coordinates,
{$$\mathcal{D}=\left\{(x,p,\dot x,\dot p):\;\; \dot p=-\frac{\partial H}{\partial x},\quad \dot x=\frac{\partial H}{\partial p}\right\}\,,$$}
so we obtain the standard Hamilton equations.

\subsection{Euler-Lagrange equations}
Let now, $\zg:\R\to M$ be a curve in $M$ (of course, $\R$ can be replaced by an open interval), and $\st\zg:\R\to\sT M$ be its tangent prolongation. It is easy to see that both curves, $\xd L\circ\st\zg$ and $\za_M\circ\st(\zl_L\circ\st\zg)$ are curves in $\sT^*\sT M$ covering $\st\zg$. Therefore, their difference makes sense and, as easily seen, takes values in the annihilator $V^0\sT M$ of the vertical subbundle $V\sT M\subset\sT\sT M$. Since $V^0\sT M\simeq\sT M\ti_M\sT^*M$, we obtain a map
$\zd L_\zg:\R\to\sT^*M$. The above map is interpreted as the external force along the trajectory. Its value at $t\in\R$ depends on the second jet $\st^2\zg(t)$ of $\gamma$ only, so defines the variation of the Lagrangian, understood as a map
\be\label{work}\zd L:\sT^2M\to\sT^*M\,,
\ee
where $\sT^2M$, the second tangent bundle, is the bundle of all second jets of curves $\R\to M$ at $0\in\R$.
The equation
\be\label{EL}\zd L_\zg=\zd L\circ\st^2\zg=0
\ee
is known as the \emph{Euler-Lagrange equation} and tells that the curve $\xd L\circ\st\zg$ corresponds \emph{via} $\za_M$ to an \emph{admissible curve} in $\sT\sT^*M$, i.e. the tangent prolongation of a curve in $\sT^*M$. Here, of course, $\st^2\zg$ is the second tangent prolongation of $\zg$ to $\sT^2M$.

\section{Mechanics on algebroids with vakonomic constraints}
The whole model of inducing dynamics out of a Lagrangian or a Hamiltonian can be repeated practically without changes when we replace $\sT M$ with a vector bundle $\zt:E\to M$ and $\zb_M$ with a map $\tilde\zP:\sT^*E^*\to\sT E^*$, associated with a linear bivector field $\zP$ on $E^*$. The fact that it reverses the direction is not really changing the picture, as we used $\za_M^{-1}$ and $\zb_M^{-1}$ to obtain the dynamics. Note that $\ze=\tilde\zP\circ\cR_E^{-1}$.
Thus, we get the diagram

$$
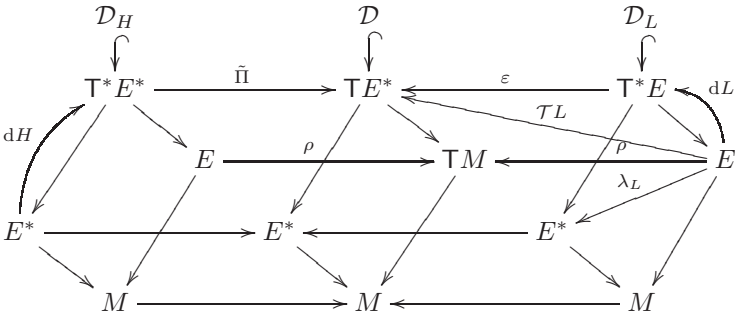
\xymatrix@C-12pt@R-10pt{
 & \mathcal{D}_H\ar@{ (->}[d] & & & {\mathcal{D}}\ar@{ (->}[d] & & & \mathcal{D}_L\ar@{ (->}[d] & \\
 & \sT^\ast E^\ast  \ar[dr] \ar[ddl]\ar[rrr]^{\tilde\Pi}
 & & & \sT E^\ast  \ar[dr] \ar[ddl]
 & & & \sT^\ast E\ar[dr]\ar[ddl]\ar[lll]_{\varepsilon} & \\
 & & E\ar[rrr]^/-10pt/{\rho}\ar[ddl]
 & & & \sT M\ar[ddl]
 & & & E \ar[ddl]\ar[lll]_/-10pt/{\rho}\ar@/_1pc/[ul]_{\xd L}\ar[dll]_{\lambda_L}\ar[ullll]_{\mathcal{T}L}  \\
 E^\ast\ar[rrr]\ar[dr]\ar@/^1pc/[uur]^{\xd H}
 & & &  E^\ast\ar[dr]
 & & & E^\ast\ar[lll]\ar[dr] & &  \\
 & M\ar[rrr] & & & M  & & & M \ar[lll]&
}$$
Here,  $\mathcal{D}_H=\xd H(E^*)\subset\sT^\ast E^\ast$ is the lagrangian submanifold associated with a Hamiltonian $H:E^\ast\longrightarrow \R$, the lagrangian submanifold $\mathcal{D}_L\subset\sT^\ast E$ is associated with a Lagrangian, and
$\mathcal{D}=\ze(\cD_L)=\mathcal{T}L(E)$ or $\mathcal{D}=\tilde\Pi(\cD_H)$, depending on the formalism used.

Starting with a Lagrangian defined on a constraint manifold $S\subset E$, we can slightly modify the above picture and get the diagram
$$
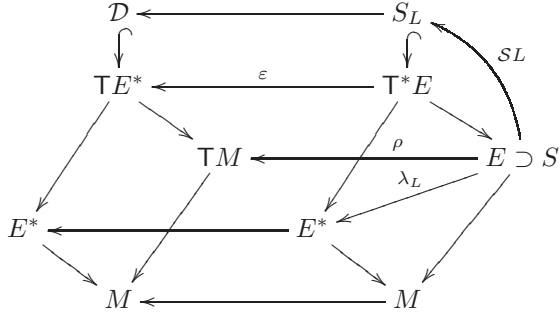
\xymatrix@C-10pt@R-10pt{
 & & {\mathcal{D}}\ar@{ (->}[d] & & & {S_L}\ar[lll]\ar@{ (->}[d] & \\
  & & \sT E^\ast  \ar[dr] \ar[ddl]
 & & &{ \sT^\ast E}\ar[dr]\ar[ddl]\ar[lll]_{\varepsilon} & \\
 & & & \sT M\ar[ddl]
 & & & {E\supset {S}} \ar[ddl]\ar[lll]_/-10pt/{\rho}\ar@/_1pc/[uul]_{\mathcal{S} L}\ar[dll]_{\lambda_L}  \\
 &  E^\ast\ar[dr]
 & & & E^\ast\ar[lll]\ar[dr] & &  \\
 & & M  & & & M \ar[lll]&
}$$
Here, $S_L$ is the lagrangian submanifold in $\sT^\ast E$ induced by the Lagrangian on the constraint $S$,
$$S_L=\{ \za_e\in\sT^\ast_eE: e\in S\text{\ and\ }\la\za_e,v_e\ran=\xd L(v_e)\text{\ for every\ }
v_e\in\sT_eS\}\,,$$
and ${\mathcal{S} L}:S\to\sT^*E$  is the corresponding relation.
The \emph{vakonomically constrained phase dynamics} is just $\mathcal{D}=\ze(S_L)\subset\sT E^*$.  We stress that, due to the fact that we are dealing with a vakonomically constrained system, relations and not just genuine smooth maps naturally appear in the formalism.

\section{Higher order Lagrangians}

The mechanics with a higher order Lagrangian $L:\sT^kQ\to\R$ is traditionally constructed as a vakonomic mechanics, thanks to the canonical embedding of of the higher tangent bundle $\sT^kQ$ into the tangent bundle $\sT\sT^{k-1}Q$ as an affine subbundle of \emph{holonomic vectors}. Thus we work with the standard Tulczyjew triple for $\sT M$, where $M=\sT^{k-1}Q$, with the presence of vakonomic constraint $\sT^kQ\subset \sT\sT^{k-1}Q$:
$$
\xymatrix@C-32pt@R-15pt{
& \sT\sT^\ast\sT^{k-1} Q \ar[dl]\ar[ddr]
&&  \sT^\ast\sT\sT^{k-1}\ar[ll] Q&& \sT^\ast\sT^k Q \ar@{-|>}[ll]\ar[dll]\ar[dd] \\
 \sT^\ast\sT^{k-1} Q\ar[rrr]\ar[ddr] & & & \sT^{k-1} Q\times_Q\sT^\ast Q \ar[ddr]+<-4.5ex,2ex>& & \\
 & & \sT\sT^{k-1} Q\ar[dl] 
& & & \sT^k Q\ar[dl]\ar@{ (->}[lll]  \\
&\sT^{k-1} Q\ar@{=}[rrr] & & &\ \sT^{k-1} Q &
}
$$

The Lagrangian function $L=L(q,\dot q,\dots,\overset{(k)}{q})$ generates the phase dynamics
$$\mathcal{D}=\left\{\, (v, p, \dot v, \dot p):\
\dot v_{i-1}=v_i,\;\; \dot p_i+p_{i-1}=\frac{\partial L}{\partial\overset{(i)}{q}}\,,{\dot p}_0=\frac{\partial L}{\partial {q}}\,,p_{k-1}=\frac{\partial L}{\partial \overset{(k)}{q}}\right\}.$$
This leads to the \emph{higher Euler-Lagrange equations} in the traditional form:
\beas
&\overset{(i)}{q}=\frac{\xd^iq}{\xd t^i}\,,\ i=1,\dots,k\,,\\
&0=\frac{\partial L}{\partial {q}}-\frac{\xd}{\xd t}\left(\frac{\partial L}{\partial {\dot q}}\right)+\cdots
+(-1)^k\frac{\xd^k}{\xd t^k}\left(\frac{\partial L}{\partial \overset{(k)}{q}}\right)\,.
\eeas
These equations can be viewed as a system of differential equations of order $k$ on $\sT^kQ$ or, which is the standard point of view, as ordinary differential equation of order $2k$ on $Q$. 


\section{Linearisation of graded bundles} The possibility of constructing mechanics on graded bundles is based on the following generalization of the embedding $\sT^kQ\hookrightarrow\sT\sT^{k-1}Q$ \cite{BGG2}.

\begin{theorem}[Bruce-Grabowska-Grabowski]
There is a canonical functor from the category of graded bundles into the category of $\mathcal{GL}$-bundles which canonically assigns, for an arbitrary graded bundle $F_k$ of degree $k$, a $\mathcal{GL}$-bundle $D(F_k)$ which is linear over $F_{k-1}$, called the \emph{linearisation of $F_k$}, together with a \emph{graded} embedding $\zi:F_k\hookrightarrow D(F_k)$ of $F_k$ as an affine subbundle of the vector bundle $D(F_k)$.
\end{theorem}

Elements of $F_k\subset D(F_k)$ may be viewed as \emph{holonomic vectors} in the linear-graded bundle $D(F_k)$. Another geometric part we need is a (Lie) algebroid structure on the vector bundle $D(F_k)\to F_{k-1}$, compatible with the second graded structure (homogeneity). We will call such $\mathcal{GL}$-bundles $D$ \emph{weighted (Lie) algebroids} and view them as abstract generalizations of the Lie algebroid $\sT\sT^{k-1}M$. Such $D$ is called a \emph{$\mathcal{VB}$-algebroid} if it is a double vector bundle.

\begin{remark}
In \cite{BGG2} a slightly more general notion of a weighted (Lie) algebroid is given for which the underlying $\mathcal{GL}$-bundle is not necessarily associated with the linearisation of a graded bundle. Furthermore, in \cite{BGG3} we discuss the corresponding groupoid objects which are named \emph{weighted Lie groupoids}. Without details, a weighted Lie groupoid is both a graded bundles and a Lie groupoid  at the same time. In a more categorical language, we have a Lie groupoid in the category of graded bundles or indeed vice versa. The basic Lie theory relating weighted Lie groupoids and weighted Lie algebroids is discussed in  \cite{BGG3}.
\end{remark}

\begin{example}{\bf (Weighted Lie algebroids out of reductions)} Let $\mathcal{G} \rightrightarrows M$ be a Lie groupoid and consider the subbundle $\sT^{k}\mathcal{G}^{\underline{s}} \subset \sT^{k}\mathcal{G}$ consisting of all higher order velocities tangent to source-leaves. The bundle
\begin{equation*}
 F_{k} = {A}^{k}(\mathcal{G}) := \left. \sT^{k}\mathcal{G}^{\underline{s}}\, \right|_{M},
 \end{equation*}
inherits graded  bundle structure of degree $k$ as a graded subbundle of $\sT^{k}\mathcal{G}$.
Of course, $A=A^1(\mathcal{G})$ can be identified with the Lie algebroid of $\mathcal{G}$.

\begin{theorem}
The linearisation of ${A}^{k}(\mathcal{G})$ is given as
$$
D({A}^{k}(\mathcal{G}) ) \simeq \{ (Y,Z) \in {A}(\mathcal{G}) \times_{M} \sT{A}^{k-1}(\mathcal{G})|\quad {\rho}(Y) = \sT {\tau}(Z)  \}\,,
$$
viewed as a vector bundle over ${A}^{k-1}(\mathcal{G})$ with respect to the obvious projection of part $Z$ onto ${{A}^{k-1}(\mathcal{G})}$, where ${\rho} : {A}(\mathcal{G}) \rightarrow \sT M$ is the standard anchor of the Lie algebroid and ${\tau}: {A}^{k-1} (\mathcal{G}) \rightarrow M$ is the obvious projection. Moreover, the above bundle is canonically a weighted Lie algebroid.
\end{theorem}
The above weighted algebroid is an example of a \emph{Lie algebroid prolongation} in the sense of Cari\~nena, {Mart\'{\i}nez}, and Popescu \cite{CM,Ma,Po}.
\end{example}

\section{Lagrangian framework for graded bundles} A weighted Lie algebroid on $D(F_k)$ gives the Tulczyjew triple
$$
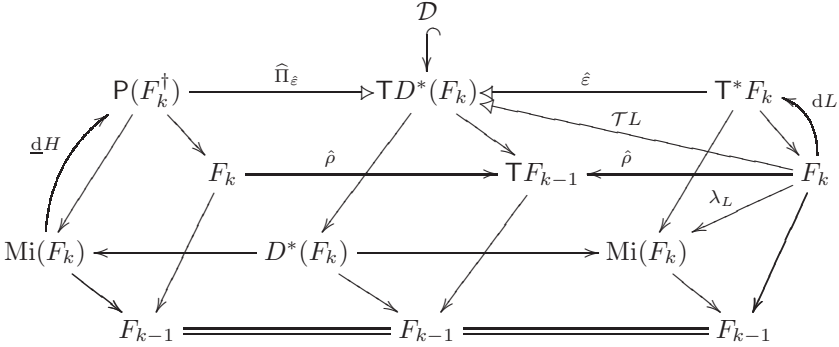
\xymatrix@C-20pt@R-10pt{
&&&& {\mathcal{D}}\ar@{ (->}[d]&&&&\\
 &\sP(F_k^\dag) \ar@{-|>}[rrr]^{\wh{\zP}_{\hat\ze}}
\ar[ddl] \ar[dr]
 &  &  & \sT D^\ast(F_k) \ar[ddl] \ar[dr]
 &  &  & \sT^\ast F_k \ar[ddl] \ar[dr]
\ar@{-|>}[lll]_{\hat\varepsilon}
 & \\
 & & F_k \ar[rrr]^/-20pt/{\hat\zr}\ar[ddl]
 & & & \sT F_{k-1}\ar[ddl]
 & & & F_k\ar[lll]_/+20pt/{\hat\zr}\ar[ddl]\ar[ddl]_{}\ar[ddl]_{}\ar@/_1pc/[ul]_{\xd L}
 \ar[dll]_{\lambda_L} \ar@{-|>}[ullll]_{\mathcal{T}L}
 \\
 \Mi(F_k)\ar[dr]\ar[dr]^{}\ar@/^1pc/[uur]^{\uxd H}
 & & & D^\ast(F_k)\ar[lll]\ar[rrr]\ar[dr]
 & & & \Mi(F_k)\ar[dr]
 & & \\
 & F_{k-1}\ar@{=}[rrr]
 & & & F_{k-1} & & & F_{k-1}\ar@{=}[lll] &
}
$$
Here, the diagram consists of relations, $\hat\ze:\sT^*F_k\rel \sT^*D(F_k)\to\sT D^*(F_k)$,  and $\Mi(F_k)$ is the so called \emph{Mironian} of $F_k$. In the classical case, $\Mi(\sT^kM)=\sT^{k-1}M\ti_M\sT^* M$. We will not discuss the Hamiltonian side of the triple and direct the reader to \cite{BGG1}. The map $\mathcal{T}L=\hat\ze\circ\xd L$ we call the \emph{Tulczyjew differential}, and $\zl_L$ the \emph{Legendre relation}.

The fact that we obtain the Euler-Lagrange equations of higher order comes from the fact that we deal with a vakonomic constraint and the additional gradation.

\begin{example} Let $\g$ be a Lie algebra and put $F_2=\g_2=\g[1]\ti\g[2]$, with coordinates $(x^i,z^j)$ on $\g_2$ and coordinates $(x^i,y^j,z^k)$ on $D(\g_2)=\g[1]\ti\g[1]\ti\g[2]$.
The embedding $\zi:\g_2\hookrightarrow D(\g_2)$ takes the form $\zi(x,z)=(x,x,z)$ and the vector bundle projection is $\zt(x,y,z)=x$. The Lie algebroid structure $\ze:\sT^*D(\g_2)\to\sT D^*(\g_2)$ reads
$$
(x,y,z,\za,\zb,\zg)\mapsto(x,\zb,\zg,z,\ad_y^*\zb,\za)\,.
$$
Given a Lagrangian $L:\g_2\to\R$, the \emph{Tulczyjew differential relation} $\mathcal{T}L:\g_2\to\sT D^*(\g_2)$ is
$$\mathcal{T}L(x,z)=\left\{\left(x,\zb,\frac{\pa L}{\pa z}(x,z),z,\ad_x^*\zb,\za\right): \za+\zb=\frac{\pa L}{\pa x}(x,z)\right\}\,.
$$
Hence, for the phase dynamics,
$$\zb=\frac{\pa L}{\pa x}(x,z)-\frac{\xd}{\xd t}\left(\frac{\pa L}{\pa z}(x,z)\right)\,.$$
This leads to the \emph{Euler-Lagrange equations} on $\g_2$:
\beas
\dot x&=&z\,,\\
\frac{\xd}{\xd t}\left(\frac{\pa L}{\pa x}(x,z)-\frac{\xd}{\xd t}\left(\frac{\pa L}{\pa z}(x,z)\right)\right)&=&\ad_x^*\left(\frac{\pa L}{\pa x}(x,z)-\frac{\xd}{\xd t}\left(\frac{\pa L}{\pa z}(x,z)\right)\right)\,.
\eeas
These equations are second order and induce the \emph{Euler-Lagrange equations} on $\g$ which are of order 3:
$$\frac{\xd}{\xd t}\left(\frac{\pa L}{\pa x}(x,\dot x)-\frac{\xd}{\xd t}\left(\frac{\pa L}{\pa z}(x,\dot x)\right)\right)=\ad_x^*\left(\frac{\pa L}{\pa x}(x,\dot x)-\frac{\xd}{\xd t}\left(\frac{\pa L}{\pa z}(x,\dot x)\right)\right)\,.
$$
For instance, the `free' Lagrangian $L(x,z)=\frac{1}{2}\sum_iI_i(z^i)^2$ induces the equations on $\g$ (here, $c^k_{ij}$ are structure constants, no summation convention):
$$I_j\dddot x^j=\sum_{i,k}c^k_{ij}I_k  x^i\ddot x^k\,.$$
The latter can be viewed as \emph{`higher Euler equations'}.
\end{example}

\section{Higher order Lagrangian mechanics on Lie algebroids}

Let us consider a general Lie groupoid $\mathcal{G}$ and a Lagrangian $L:{A}^{k}\to\R$ on $A^k={A}^{k}(\mathcal{G})$.
We will refer to such systems as a \emph{k-th order Lagrangian system on the Lie algebroid} ${A}(\mathcal{G})$. The relevant diagram here is
$$
\xymatrix@C-20pt@R-5pt{
{\mathcal{D}\subset\hskip-.3cm}&\sT D^{*}({A}^{k}(\mathcal{G}))\ar[dd]\ar[dr] && &\sT^{*} D({A}^{k}(\mathcal{G}))\ar[lll]_\ze\ar[dll]  && & \sT^{*}{A}^{k}(\mathcal{G})\ar@{-|>}[lll]_{\ r}\ar[dd]    \\
 && D^{*}({A}^{k}(\mathcal{G})) &&& &&\\
            & \sT {A}(\mathcal{G})  && & D({A}^{k}(\mathcal{G}))\ar[lll]_\zr  & && {A}^{k}(\mathcal{G})\ar[lll]_\zi\ar@/_1pc/[uu]_{\xd L}\ar@{-|>}[ulllll]_{\zl_L}
}
$$
Here, $D({A}^{k}(\mathcal{G}) )$ is the corresponding Lie algebroid prolongation, $\mathcal{D}=\ze\circ r\circ\xd L({A}^{k}(\mathcal{G}))$, and $\zl_L$ is the \emph{Legendre relation}.
Note that we actually deal with reductions: if $\mathcal{G}$ is a Lie group, then
$${A}^k(\mathcal{G})=\sT^k(\mathcal{G})/\mathcal{G}\quad \text{and}\quad D({A}^{k}(\mathcal{G}) )=\sT\sT^{k-1}(\mathcal{G})/\mathcal{G}\,.$$
For instance, using $x^A$ as base coordinates, and $y^a_i$ as fibre coordinates of degree $i=1,\dots, k$ in $A^{k}$, extended by the appropriate momenta $\zp^j_b$ of degree $j=1,\dots, k$ in $D^*(A^k)$, we get the equations for the Legendre relation in the form \emph{(no Lie algebroid structure is relevant)}:

\begin{tabular}{l}
                                                               $\displaystyle k\pi_{a}^{1} = \frac{\partial L}{\partial {y}^{a}_{k}}$,\\
                                                               $\displaystyle (k-1)\pi_{b}^{2} = \frac{\partial L}{\partial {y}^{b}_{k-1}} - \frac{1}{k}\frac{d}{dt}\left(  \frac{\partial L}{\partial {y}^{b}_{k}} \right)$,\\
                                                               $\displaystyle \vdots $\\
                                                               $ \displaystyle\pi_{d}^{k} = \frac{\partial L}{\partial {y}_{1}^{d}} - \frac{1}{2!}\frac{d}{dt}\left(\frac{\partial L}{\partial {y}_{2}^{d}}\right) + \frac{1}{3!}\frac{d^{2}}{dt^{2}}\left(\frac{\partial L}{\partial {y}_{3}^{d}}\right)- \cdots $\\
                                                               $\displaystyle + (-1)^{k} \frac{1}{(k-1)!} \frac{d^{k-2}}{dt^{k-2}}\left(\frac{\partial L}{\partial {y}_{k-1}^{d}}\right) - (-1)^{k} \frac{1}{k!}\frac{d^{k-1}}{dt^{k-1}}\left(\frac{\partial L}{\partial {y}_{k}^{d}}\right)$,
                                                              \end{tabular}

\medskip\noindent which we recognise as the \emph{Jacobi--Ostrogradski momenta}.
The remaining equation for the dynamics makes use of the Lie algebroid structure and reads
$$\frac{d}{dt} \pi^{k}_{a} = \rho_{a}^{A}({x}) \frac{\partial L}{\partial {x}^{A}} + {y}_{1}^{b}C_{ba}^{c}({x})\pi_{c}^{k}\,,$$
where $\rho_{a}^{A}$ and $C_{ba}^{c}$ are structure functions of the Lie algebroid $A={A}(\mathcal{G})$.

The above equation can then  be rewritten as
\beas
     &\rho_{a}^{A}({x}) \frac{\partial L}{\partial {x}^{A}}=&\\   &\left(\delta^{c}_{a} \frac{d}{dt} - {y}_{1}^{b}C_{ba}^{c}({x}) \right)\left(  \frac{\partial L}{\partial {y}_{1}^{c}} - \frac{1}{2!}\frac{d}{dt}\left(\frac{\partial L}{\partial {y}_{2}^{c}}\right) \cdots  {-} (-1)^{k}\frac{1}{k!}\frac{d^{k-1}}{dt^{k-1}}\left(\frac{\partial L}{\partial {y}_{k}^{c}}\right)\right)&\,,
\eeas
which we define to be the \emph{k-th order Euler--Lagrange equations} on ${A}(\mathcal{G})$.

The above higher order algebroid E-L equations are in complete agrement  with the ones obtained by {Colombo \& de Diego} \cite{CD}, {J\'{o}\'{z}wikowski \&   Rotkiewicz} \cite{JR1,JR2}, as well as {Mart\'{\i}nez} \cite{Ma1}. We clearly recover the standard higher Euler--Lagrange equations on $\sT^{k}M$ as a particular example.

\begin{example} {\bf (The tip of a javelin)} For instance, let $L$ be the Lagrangian governing the motion of the tip of a javelin
defined on $\sT^2\R^3$,
$$
L(x,y,z)=\frac12\left(\sum_{i=1}^3(y^i)^2-(z^i)^2\right) \,.
$$
We can understand $G=\R^3$ here as a commutative Lie group, and since $L$ is $G$-invariant, we get immediately the reduction to the graded bundle $\R^3[1]\ti\R^3[2]$.
The Euler-Lagrange equations on $\sT^2\R^3$,
$$\frac{\xd}{\xd t}\left(\frac{\partial L}{\partial y^i} - \frac{1}{2}\frac{\xd}{\xd t}\left(\frac{\partial L}{\partial z^i}\right)\right)=0\,,
$$
give in this case
$$
\frac{\xd y^i}{\xd t}=\frac{1}{2}\frac{\xd^2 z^i}{\xd t^2}\,,
$$
so the Euler-Lagrange equation on $\R^3$ reads
$$
\frac{\xd^2 x^i}{\xd t^2}=\frac{1}{2}\frac{\xd^4 x^i}{\xd t^4}\,.
$$
\end{example}
\section{Geometric mechanics of strings}
Another example of a Tulczyjew triple constructed from double graded bundles and morphisms is
the one in which the dynamics lives in $\we^n\sT\we^n\sT^*M$ (see \cite{GGU}).
The justification is the following. We want to build a framework for higher dimensional objects, being motivated by the study of dynamics of one-dimensional non-parameterized  objects  (strings).

The \emph{motion} of a system  will be given by an $n$-dimensional submanifold in the manifold $M$ (``space-time'').  Therefore, an infinitesimal piece of the motion is the first jet of the submanifold.
However, this model leads to essential complications even in one-dimensional case (relativistic particle).  For instance, the infinitesimal action (Lagrangian) is not a function on first jets, but a section of certain line bundle over the first-jet manifold, a `dual' of the bundle of ``first jets with volumes''.
Therefore we will take the compromise: use for the space of infinitesimal pieces of motions the space of simple $n$-vectors, which represent first jets of $n$-dimensional submanifolds together with an infinitesimal volume.  It is technically convenient to extend this space to all $n$-vectors, i.e. to  the vector bundle $\wedge^n\sT M$ of $n$-vectors on $M$.
In this way we get the following principles:
\begin{itemize}
\item A \emph{Lagrangian} $L$ is a function on infinitesimal motions, $L:\wedge^n\sT M\to\R$.
 If $L$  is positive homogeneous, the action functional does not depend on the parametrization of the submanifold and the corresponding Hamiltonian (if it exists) is a function on the dual vector bundle  $\wedge^n\sT^\* M$ (the phase space).
\item The \emph{dynamics} should be an equation (in general, implicit) for $n$-dimensional submanifolds in the phase space, i.e.
    $${\cD}\subset\wedge^n \sT \wedge^n\sT^\* M\,.$$
\item A submanifold $N$ in the phase space $\wedge^n\sT^\* M$ is a \emph{solution} of $\cD$ if and only if its tangent space $\sT_\za N$ at $\za\in\wedge^n\sT^\* M$ is represented by a $n$-vector from ${\cD}_\za$. If we use a parametrization, then the tangent $n$-vectors associated with this parametrization must belong to $\cD$.
\end{itemize}
For simplicity, in what follows we will consider the `string case' $n=2$, but the constructions remain valid for arbitrary $n$. We will use canonical coordinates $(x^\zr,\dot x^{\zm\zn})$ and $(x^\zr,p_{\zm\zn})$ on $\we^2\sT M$ and $\we^2\sT^*M$ (with the convention
    $\dot x^{\zm\zn}=-\dot x^{\zn\zm}$, $p_{\zm\zn}=-p_{\zn\zm})$, respectively, representing the decomposition of bivectors:
    $$\dot x^{\zm\zn}\pa_{x^\zm}\we\pa_{x^\zn}\in\we^2\sT M\,,\quad p_{\zm\zn}\xd x^\zm\we\xd x^\zn\in\we^2\sT^*M\,.$$
Using the canonical multisymplectic structure on $\wedge^2\sT^\*M$, we get the following \emph{Tulczyjew triple} for multivector bundles, consisting of \emph{double graded bundle morphisms}:
$$
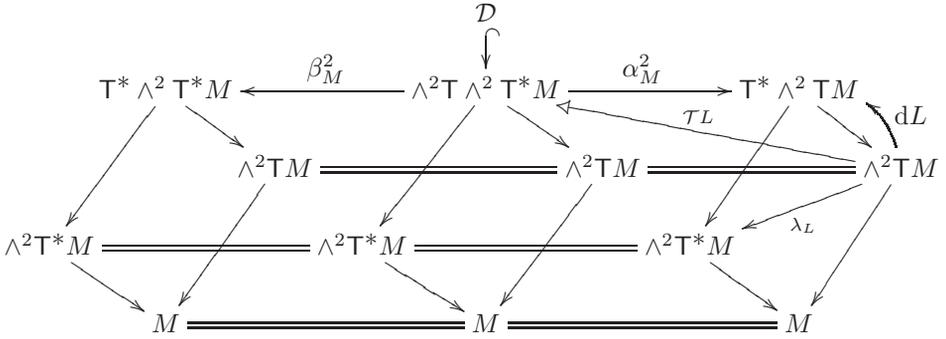
{\xymatrix@R-3mm @C-10mm{
    &&&&  \mathcal{D}\ar@{ (->}[d]&&&\\
    &  \sT^\*\wedge^2\sT^\*M  \ar[ldd]_*{} \ar[rd]^*{} & & & \wedge^2 \sT \wedge^2\sT^\* M \ar[rrr]^*{{\za}^2_M}
    \ar[lll]_*{\zb^2_M} \ar[ldd]^*{} \ar[rd]^*{}& & & {\sT^\*\wedge^2\sT M} \ar[ldd]^*{} \ar[rd]^*{} & \cr
    & & \wedge^2\sT M \ar[ldd]^*{} & & & \wedge^2\sT M  \ar[ldd]^*{} \ar@{=}[lll]^*{} \ar@{=}[rrr]^*{} & & & \wedge^2\sT M \ar[ldd]^*{}\ar@/_1pc/[ul]_(.3)*{\xd L\ }\ar@{-|>}[ullll]_{\mathcal{T}L}\ar[lld]^{\zl_L} \cr
    \wedge^2\sT^\* M  \ar[rd]^*{}  & & & \wedge^2\sT^\* M \ar@{=}[rrr]^*{} \ar@{=}[lll]^*{} \ar[rd]^*{}& & & \wedge^2\sT^\* M
    \ar[rd]^*{} & & \cr
    & M  & & &  M \ar@{=}[rrr]^*{} \ar@{=}[lll]^*{} & & & M & }}\,.
$$
The way of obtaining the implicit phase dynamics $\mathcal{D}$, as a submanifold of $\wedge^2 \sT \wedge^2\sT^\* M$, from a Lagrangian $L:\wedge^2\sT M\to\R$ (or from a Hamiltonian $H:\wedge^2\sT^\* M\to\R$) is now standard: $\mathcal{D}=\mathcal{T}L(\we^2\sT M)$.

\subsection{The Euler-Lagrange equations}
To define \emph{Euler-Lagrange equations} for strings, consider
a surface
$S:\R^2\ni(t,s)\mapsto (x^\zs(t,s))$
in $M$ and its bi-tangent prolongation
$$\we^2\st\, S:\R^2\to\we^2\sT M\,,\quad\we^2\st\, S=\st_tS\we\st_sS\,.
$$
On the diagram:
$$
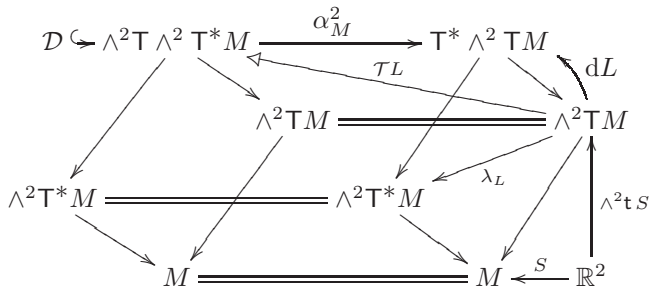
{\xymatrix@R-3mm @C-10mm{
    {\mathcal{D}}\ar@{ (->}[r] & \wedge^2 \sT \wedge^2\sT^\* M \ar[rrr]^*{{\za}^2_M}
     \ar[ldd]^*{} \ar[rd]^*{}& & & {\sT^\*\wedge^2\sT M} \ar[ldd]^*{} \ar[rd]^*{} & \cr
     && \wedge^2\sT M  \ar[ldd]^*{} \ar@{=}[rrr]^*{} & & & \wedge^2\sT M \ar[ldd]^*{}\ar@/_1pc/[ul]_(.3)*{\xd L\ }\ar@{-|>}[ullll]_{\mathcal{T}L}\ar[lld]^{\zl_L} \cr
     \wedge^2\sT^\* M \ar@{=}[rrr]^*{}  \ar[rd]^*{}& & & \wedge^2\sT^\* M
    \ar[rd]^*{} & & \cr
     &  M \ar@{=}[rrr]^*{}  & & & M & \R^2\ar[l]_{ S}\ar[uu]_{\we^2\st\, S}}}\,.
$$
It is easy to see that both parameterized surfaces, $\xd L\circ\we^2\st\, S$ and $\za^2_M\circ\we^2\st(\zl_L\circ\we^2\st\, S)$ in $\sT^*\we^2\sT M$ cover $\we^2\st\, S$. Therefore, their difference makes sense and, as easily seen, takes values in the annihilator $V^0\we^2\sT M$ of the vertical subbundle $V\we^2\sT M\subset\sT\we^2\sT M$. Since $V^0\we^2\sT M\simeq\we^2\sT M\ti_M\sT^*M$, we obtain a map
$\zd L_S:\R^2\to\sT^*M$. The above map is interpreted as external forces along the string trajectory $S$. Its value at $(t,s)$ depends on the second jet $\sj^2 S(t,s)$ of $S$ only, so defines the variation of the Lagrangian understood as a map
\be\label{works}\zd L:\sJ^2_0(\R^2,M)\to\sT^*M\,,
\ee
where $\sJ^2_0(\R^2,M)$ is the bundle of all second jets of maps $\R^2\to M$ at $0\in \R^2$.
The equation
\be\label{ELs}\zd L_S=0
\ee
we will call the \emph{Euler-Lagrange equation}. It tells that the surface $\xd L\circ\we^2\st\, S$ corresponds \emph{via} $\za^2_M$ to an \emph{admissible surface} in $\we^2\sT\we^2\sT^*M$, i.e. the bi-tangent prolongation of a parameterized surface in $\we^2\sT^*M$.

A surface $S:(t,s)\mapsto (x^\zs(t,s))$ in $M$ satisfies the Euler-Lagrange equations if the image by $\xd L$  of its prolongation to $\we^2\sT M$,
$$
(t,s)\mapsto \left(x^\zs(t,s),\dot x^{\zm\zn}=\frac{\pa x^\zm}{\pa t}\frac{\pa x^\zn}{\pa s}-\frac{\pa x^\zm}{\pa s}\frac{\pa x^\zn}{\pa t}\right)\,,
$$
is $\za_M^2$-related to an admissible surface, i.e. the prolongation of a surface, living in the phase space $\we^2\sT^\ast M$, to $\we^2\sT\we^2\sT^\ast M$.

In coordinates, the Euler-Lagrange equations read

\beas\dot x^{\zm\zn}&=&\frac{\pa x^\zm}{\pa t}\frac{\pa x^\zn}{\pa s}-\frac{\pa x^\zm}{\pa s}\frac{\pa x^\zn}{\pa t}\,,\\
\frac{\pa L}{\pa x^\zs}&=&\frac{\pa x^\zm}{\pa t}\frac{\pa}{\pa s}\left(\frac{\pa L}{\pa \dot x^{\zm\zs}}(t,s)\right)-\frac{\pa x^\zm}{\pa s}\frac{\pa}{\pa t}\left(\frac{\pa L}{\pa \dot x^{\zm\zs}}(t,s)\right)\,.
\eeas

\medskip
\begin{example} {\bf (Plateau problem)} In particular, if $M=\R^3=\{(x^1=x,x^2=y,x^3=z)\}$ with the Euclidean metric, the canonically induced `free' Lagrangian on $\we^2\sT M$ reads
$$
L(x^\zm,\dot x^{\zk\zl})=\sqrt{\sum_{\zk,\zl}\left(\dot x^{\zk\zl}\right)^2}\,.
$$
The Euler-Lagrange equation for surfaces being graphs $(x,y)\mapsto (x,y,z(x,y))$ provides the well-known equation for \emph{minimal surfaces}, found already by Lagrange:
$$
\frac{\pa}{\pa x}\left(\frac{z_x}{\sqrt{1+z_x^2+z_y^2}}\right)+\frac{\pa}{\pa y}\left(\frac{z_y}{\sqrt{1+z_x^2+z_y^2}}\right)=0\,.
$$
In another form,
$$(
1+z_x^2)z_{yy}-2z_xz_yz_{xy}+(1+z_y^2)z_{xx}=0\,.
$$
Starting with a Lorentz metric, we can obtain analogously the Euler--Lagrange equations for the \emph{Nambu-Goto Lagrangian}.
\end{example}

\section{Concluding remarks}
We hope that the reader now appreciates that graded bundles and multi-graded bundles are interesting objects from a geometric perspective, but moreover that they are potentially very important in geometric mechanics.  We have only `scratched the surface' here with the applications of graded bundles in geometric mechanics and expect further results to follow.

The name of this volume suggest one line of further investigation: that is to develop jets and field theories within the generale framework presented here. A little more specifically, the notion of a \emph{Lie algebroid valued jet} was developed by Mart\'{\i}nez \cite{Martinez:2005} and it is natural to wonder what further structure come available when passing to weighted Lie algebroids. From there one may be able to develop a framework for higher order field theory using the Tulczyjew triple approach, modifying the constructions of \cite{Grabowska:2012,Grabowska:2015} as needed. This is work to be done in the near future.

In conclusion, the  formalism of (n-tuple) graded bundles offers a clear and powerful way to view many constructions in differential geometry, most notably
iterations of (higher) tangent bundles \& cotangent bundles, as well as multivector bundles. Phrasing non-negatively graded geometry in terms of homogeneity structures  can offer solutions to questions that would not otherwise be so readily obtainable. As an example we offer the problem of integrating weighted Lie algebroids as presented in \cite{BGG1}.

\subsection*{Acknowledgments} Research funded by the Polish National Science Centre grant under the contract number DEC-2012/06/A/ST1/00256


\begin{thebibliography}{99}

\bibitem{BGG3} A.~J.~Bruce, K.~Grabowska, J.~Grabowski, \emph{Graded bundles in the category of Lie groupoids}, arXiv:1502.06092.

\bibitem{BGG1} A.~J.~Bruce, K.~Grabowska, J.~Grabowski, \emph{Higher order mechanics on graded bundles}, J. Phys. A {48} (2015), 205203--205235.

\bibitem{BGG2} A.~J.~Bruce, K.~Grabowska, J.~Grabowski, \emph{Linear duals of graded bundles and higher analogues of (Lie) algebroids}, arXiv:1409.0439.

\bibitem{CM}
J.~F.~Cari\~{n}ena and E.~Mart\'{\i}nez,
\emph{Lie algebroid generalization of geometric mechanics},
In \textsl{Lie algebroids and related topics in differential geometry} (Warsaw, 2000), 201-215,
Banach Center Publ. {54}, Polish Acad. Sci. Inst. Math., Warsaw, 2001.

\bibitem{CD}
L.~Colombo and D.~M.~de~Diego,
\emph{Higher-order variational problems on lie groups and optimal control applications},
J.~Geom.~Mech. {6} (2014), 451--478.

\bibitem{GGU} K.~Grabowska, J.~Grabowski, P.~Urba\'nski,  \emph{Geometry of Lagrangian and Hamiltonian formalisms in the dynamics of strings}, J.~Geom.~Mech. {6} (2014), 503--526

\bibitem{Grabowska:2006}
{K.~Grabowska,  J.~~Grabowski and  P. Urba\'{n}ski,}
\emph{Geometrical mechanics on algebroids,}
{Int. J. Geom. Methods Mod. Phys. 3 (2006), 559--575.}

\bibitem{Grabowska:2012}
K.~Grabowska,
\emph{The Tulczyjew triple for classical fields},
J. Phys. A, 45 (2012), 145207--145242.

\bibitem{Grabowska:2015}
K.~Grabowska and L.~Vitagliano,
\emph{Tulczyjew Triples in Higher Derivative Field Theory},
 J. Geom. Mec. 7 (2015) 1--33.


\bibitem{GR0}
J.~Grabowski and M.~Rotkiewicz,
\emph{Higher vector bundles and multi-graded symplectic manifolds}, J.~Geom.~Phys. {59} (2009), 1285--1305.

\bibitem{GR} J.~Grabowski and M.~Rotkiewicz, \emph{Graded bundles and homogeneity structures}, J.~ Geom.~Phys. {62} (2012), 21--36.

\bibitem{GU3}
J.~Grabowski and P.~Urba\'nski,
\emph{Lie algebroids and Poisson-Nijenhuis structures},
Rep.~Math.~ Phys. {40} (1997), 195--208.

\bibitem{GU1}
J.~Grabowski and P.~Urba\'{n}ski,
\emph{Algebroids - general differential calculi on vector bundles},
J.~Geom.~Phys. {31} (1999),  111--141.

\bibitem{JR1}
M.~J\'{o}\'{z}wikowski and  M.~Rotkiewicz,
\emph{Prototypes of higher algebroids with applications to variational calculus},
\texttt{arXiv:1306.3379v2 [math.DG]} (2014).

\bibitem{JR2}
M.~J\'{o}\'{z}wikowski and  M.~Rotkiewicz,
\emph{Models for higher algebroids},
J.~Geom.~Mech. {7} (2015), 317--359.

\bibitem{KU} K.~Konieczna and P.~Urba\'nski,
\emph{Double vector bundles and duality},
Arch.~Math. (Brno) {35} (1999), 59--95.



\bibitem{Ma}
E.~Mart\'{\i}nez,
\emph{Lagrangian Mechanics on Lie algebroids},
Acta Appl.~Math. {67} (2001), 295--320.

\bibitem{Martinez:2005}
E.~Mart\'{\i}nez,
\emph{Classical field theory on Lie algebroids: variational aspects},
J. Phys. A: Math. Gen. 38 (2005)  7145.

\bibitem{Ma1}
E.~Mart\'{\i}nez,
\emph{Higher-order variational calculus on Lie algebroids},
J.~Geom.~Mech. {7} (2015), 81--108.

\bibitem{Mc}
K.~C.~H.~Mackenzie and P.~Xu,
\emph{Lie bialgebroids and Poisson groupoids},
Duke Math.~J. {73} (1994), 415--452.

\bibitem{Po}
M.~Popescu and P.~Popescu,
\emph{Geometric objects defined by almost Lie structures},
In {\sl Lie algebroids and related topics in differential geometry} (Warsaw, 2000),  217--233, {Banach Center Publications} {54} (2001), Polish Acad. Sci., Warsaw, 2001.

\bibitem{P} J.~Pradines, 
\emph{Repr\'{e}sentation des jets non holonomes par des morphismes vectoriels doubles soud\'{e}s}, C.R. Acad. Sci. Paris, s\'{e}rie A {278} (1974), 1523--1526.

\bibitem{Roytenberg:2001}
D.~Roytenberg,
\emph {On the structure of graded symplectic supermanifolds and {C}ourant algebroids}.
In {\sl Quantization, Poisson Brackets and Beyond}, volume 315 of
  Contemp. Math., Amer. Math. Soc., Providence, RI, 2002.

\bibitem{Severa:2005}
P.~\v{S}evera,
\emph{Some title containing the words ``homotopy''  and ``symplectic'', e.g. this one, }
{Travaux math\'{e}matiques, Univ. Luxemb. {16} (2005), 121-137.}

\bibitem{Tulczyjew:1976} W.~M.~Tulczyjew, 
\emph{Les sous-varietes Lagrangiennes et la Dynamique Hamiltonienne}, C.~R.~Acad.~Sc.~Paris {\bf 283} (1976), 15--18.

\bibitem{Tulczyjew:1976a} W.~M.~Tulczyjew, 
\emph{Les sous-varietes Lagrangiennes et la Dynamique Lagrangienne}, C.~R.~Acad.~Sc.~Paris {\bf 283} (1976) 675--678.

\bibitem{Tulczyjew:1989} W.~M.~Tulczyjew, 
\emph{Geometric Formulation of Physical Theories. Statics and Dynamics of Mechanical Systems},
 Monographs and Textbooks in Physical Science. Lecture Notes 11, Bibliopolis, Naples, 1989.

\bibitem{Ur} P.~Urba\'nski,
\emph{Double vector bundles in classical mechanics},
Rend.~Sem.~Matem.~Torino {54} (1996), 405--421.

\bibitem{Voronov:2001qf}
Th.Th.~Voronov,
\emph {Graded manifolds and {D}rinfeld doubles for {L}ie bialgebroids},
In { \sl Quantization, Poisson Brackets and Beyond}, volume 315 of
  Contemp. Math., pages 131--168. Amer. Math. Soc., Providence, RI, 2002.

\bibitem{Weinstein:1996}
A.~Weinstein,
\emph{Lagrangian Mechanics and groupoids,}
{Fields Inst. Comm., 7 (1996), 207-231.}

\end{thebibliography}
\end{document}